\documentclass[a4paper,11pt]{article}
\usepackage{pos}
\renewcommand{\logo}{\relax}

\newcommand{\Z}{\mathbb{Z}}
\newcommand{\toCP}{\xrightarrow{CP}}

\newcommand{\mmmatrix}[9]{ \left(\!\! \begin{array}{ccc}#1 & #2 & #3\\ #4 & #5 & #6\\ #7 & #8 & #9\\ \end{array}\!\!\right) }

\newcommand{\triplet}[3]{ \left(\! \begin{array}{c}#1 \\ #2 \\ #3 \end{array}\!\right) }
\newcommand{\lr}[1]{ \langle #1 \rangle}

\usepackage{tikz}

\title{Confronting CP symmetry of order 4 with experimental data}

\author*{Igor P. Ivanov}
\author{Duanyang Zhao}

\affiliation{School of Physics and Astronomy, Sun Yat-sen University,\\
	Daxue road 2, Tangjiawan, 519082 Zhuhai, China}

\emailAdd{ivanov@mail.sysu.edu.cn}
\emailAdd{zhaody8@mail2.sysu.edu.cn}

\abstract{CP4 3HDM is a three-Higgs-doublet model based on a $CP$ symmetry of order 4 (CP4).
It is the minimal model incorporating CP4 without leading to accidental symmetries 
or running into immediate conflict with experiment. 
Imposing CP4 on the lagrangian induces remarkably tight connections between the scalar and Yukawa sectors,
including the unavoidable tree-level flavor-changing neutral couplings (FCNC).
Here, we explore whether it is at all possible in the CP4 3HDM to suppressed FCNC to a level compatible 
with the neutral meson oscillation constraints.
We express the FCNC matrices in terms of physical quark observables and quark rotation parameters,
and scan the Yukawa parameter space using the quark masses and mixing parameters as input.
With this procedure, we find that only two out of the eight possible CP4 Yukawa sectors
are compatible with the $K$, $B$, $B_s$ and, in particular, $D$-meson oscillation constraints.
The results clearly indicate a way how to construct phenomenologically viable benchmark CP4 3HDMs.
}

\FullConference{Corfu Summer Institute 2023 "School and Workshops on Elementary Particle Physics and Gravity" (CORFU2023)\\
 23 April - 6 May , and 27 August - 1 October, 2023\\
Corfu, Greece\\}

\begin{document}
\maketitle

\section{The curious case of CP4 3HDM}


Multi-Higgs-doublet models are an attractive framework to build models beyond the Standard Model (bSM).
The simple idea of generations applied to the Higgs sector, 
together with very few assumptions, can lead to numerous phenomenological and cosmological consequences.
The two-Higgs-doublet model (2HDM) proposed by T.~D.~Lee half a century ago \cite{Lee:1973iz}
has become a standard playground for collider searches of additional Higgs bosons \cite{Branco:2011iw,LHCHiggsCrossSectionWorkingGroup:2016ypw}.
The three-Higgs-doublet model (3HDM), first proposed by S.~Weinberg in 1976 \cite{Weinberg:1976hu} 
as a means of combining $CP$ violation with natural flavor conservation,
was actively explored in 1980's and flourished again in the past decade,
see \cite{Ivanov:2017dad} for a brief historical overview.
There is already an extensive literature on models using four or more Higgs doublets, see \cite{Shao:2023oxt} for a relevant literature review.
In short, the multi-Higgs-doublet framework remains an actively explored and phenomenologically attractive option,
with new phenomenological opportunities still being explored.

In most cases, multi-Higgs models are shaped by additional global symmetries which act on scalars and fermions.
Various options for the symmetry groups and representation choices have been used. 
For example, the 2HDM with a softly broken $\Z_2$ symmetry \cite{Branco:2011iw}
allows one to implement the natural flavor conservation principle~\cite{Glashow:1976nt,Paschos:1976ay,Peccei:1977hh}
and, as a result, to avoid the tree-level Higgs-mediated flavor changing neutral couplings (FCNC).
Eliminating FCNCs altogether is not compulsory, though. Certain amount of tree-level FCNCs
can be tolerated if they are sufficiently suppressed \cite{Sher:2022aaa}.
Remarkably, in certain symmetry-based models, 
such as the famous Branco-Grimus-Lavoura (BGL) model \cite{Branco:1996bq} 
and its generalizations \cite{Botella:2014ska,Botella:2015hoa,Alves:2017xmk},
this suppression is naturally controlled by the small elements of the Cabibbo-Kobayashi-Maskawa (CKM) matrix.

The 2HDMs, however, offer a very limited choice of global symmetry groups
\cite{Ivanov:2006yq,Nishi:2006tg,Ivanov:2007de,Maniatis:2007de,Ferreira:2009wh}.
3HDMs can accommodate many more symmetry options for model building \cite{Ivanov:2012ry,Ivanov:2012fp,Darvishi:2019dbh},
including several non-abelian finite groups, as well as novel options for $CP$ violation.
What is less known is that the 3HDMs can also accommodate a new form of $CP$ symmetry,
which is physically distinguishable from the traditional $CP$ \cite{Haber:2018iwr}.
It was noticed long ago that the action of discrete symmetries, such as $CP$,
on quantum fields is not uniquely defined \cite{feinberg-weinberg,Lee:1966ik,Branco:1999fs,weinberg-vol1}.
If a model contains several fields with identical gauge quantum numbers,
this freedom in defining $CP$ is further enhanced:
one can consider a general $CP$ transformation (GCP) which not only maps the fields to their conjugates
but also mixes them \cite{Ecker:1987qp,Grimus:1995zi}:
\begin{equation}
	\phi_i({\bf r}, t) \toCP {\cal CP}\,\phi_i({\bf r}, t)\, ({\cal CP})^{-1} = X_{ij}\phi_j^*(-{\bf r}, t), \quad X_{ij} \in U(N)\,.
	\label{GCP}
\end{equation}
Here, we showed the GCP action on the complex scalar fields $\phi_i$, $i = 1, \dots, N$.
The conventional $CP$ choice $X_{ij} = \delta_{ij}$ is just one of many possible
choices and is basis-dependent.
If a model does not respect the conventional $CP$ but is invariant under a GCP with a suitable matrix $X$,
then the model is explicitly $CP$-conserving \cite{Branco:1999fs}.

The presence of the matrix $X$ has consequences.
Applying the GCP twice leads to a Higgs family transformation, $\phi_i \mapsto (XX^*)_{ij}\phi_j$,
which may be non-trivial.
It may happen that one arrives at the identity transformation only when the GCP transformation is applied $k$ times;
by definition, such a $CP$ symmetry has order $k$.
The conventional $CP$ is of order 2; the next option is a $CP$ symmetry of order 4, denoted CP4.
Since the order of a transformation is basis-invariant, a model based on a CP4
represents a physically distinct $CP$-invariant model which cannot be achieved with the conventional $CP$.

Within the 2HDM, imposing CP4 on the scalar sector always leads to the usual $CP$ \cite{Ferreira:2009wh}.
In order to implement CP4 and avoid any conventional $CP$, one needs to pass to three Higgs doublets.
This model, dubbed CP4 3HDM, was constructed in \cite{Ivanov:2015mwl} building upon results of \cite{Ivanov:2011ae}
and was found to possess remarkable features which sometimes defy intuition \cite{Ivanov:2015mwl,Aranda:2016qmp,Haber:2018iwr}.
If CP4 remains unbroken at the minimum of the Higgs potential, it can protect the scalar dark matter candidates against decay
\cite{Ivanov:2018srm} and may be used to generate radiative neutrino masses \cite{Ivanov:2017bdx}.
Multi-Higgs-doublet models based on even higher order $CP$ symmetries were constructed in \cite{Ivanov:2018qni}.

CP4 symmetry can also be extended to the Yukawa sector leading to very particular patterns of the Yukawa matrices \cite{Ferreira:2017tvy}.
In order to avoid mass degenerate quarks, the CP4 symmetry of the model must be spontaneously broken.
Then, it turns out that the Yukawa sector contains enough free parameters to accommodate
the experimentally measured quark masses and mixing, as well as the appropriate amount of $CP$ violation.
Tree-level Higgs-mediated FCNCs are unavoidable in the CP4 3HDM and must be checked.
The scalar alignment assumption used in \cite{Ferreira:2017tvy} guarantees that
the SM-like Higgs $h_{SM}$ does not change the quark flavor.
But the other neutral scalars certainly do.
The scan performed in \cite{Ferreira:2017tvy} produced parameter space points
which satisfied the kaon and $B$-meson oscillation parameters.
However the FCNC processes in the up-quark sector, such as $D$-meson oscillations or flavor-changing top-quark
processes, were not included in \cite{Ferreira:2017tvy}.

A few years later, Ref.~\cite{Ivanov:2021pnr} revealed that the charged Higgs boson couplings
to the top quark ruled out the vast majority of the parameter space points believed viable in \cite{Ferreira:2017tvy}.
Almost all examples found in \cite{Ferreira:2017tvy}
contained one or two charged Higgs bosons lighter than the top quark,
opening up non-standard top decay channels, which entered in conflict with the LHC Run 2 data
or with the total top width measurements.

Clearly, one needs to repeat the parameter space scan including the new constraints.
However, an additional downside of \cite{Ferreira:2017tvy} was that only
a very small fraction of the random scan points passed all the flavor constraints.
This was due to the wide-spread but intrinsically inefficient scanning method, which used the parameters of the lagrangian as input.
In this method, one begins with a random point in the parameter space, finds quark masses and mixing parameters way off
their experimental values, and then iteratively seeks for a suitable combination of the parameters
which would lead to the measured quark masses and mixing values.
It would be much more efficient to use the physical observables --- quark masses and mixing parameters ---
as input, and then to proceed directly to computing the FCNC processes. 

Here, we report the results of the recent work \cite{Zhao:2023hws}, in which we constructed 
this scanning procedure and, using it, dramatically limited the number of viable 
CP4 3HDM Yukawa sectors. 


\section{3HDM with $CP$-symmetry of order 4}\label{section-CP4}

\subsection{The scalar sector of CP4 3HDM}

The 3HDMs make use of three Higgs doublets $\phi_i$, $i = 1,2,3$ with identical quantum numbers.
CP4 acts on Higgs doublets by conjugation accompanied with a rotation in the doublet space.
Following \cite{Ivanov:2015mwl,Ferreira:2017tvy}, we use the following form of the CP4:
\begin{equation}
	\phi_i \toCP X_{ij} \phi_j^*\,,\quad
	X =  \left(\begin{array}{ccc}
		1 & 0 & 0 \\
		0 & 0 & i  \\
		0 & -i & 0
	\end{array}\right)\,.
	\label{CP4-def}
\end{equation}
Applying this transformation twice leads to a non-trivial transformation in the space of doublets:
$\phi_{1} \mapsto \phi_{1}$, $\phi_{2,3} \mapsto -\phi_{2,3}$.
In order to get the identity transformation, we must apply CP4 four times, hence the order-4 transformation.
Any CP-type transformation of order 4 acting in the space of three complex fields
can always be presented in the form \eqref{CP4-def} by a suitable basis change \cite{weinberg-vol1}.

The most general renormalizable 3HDM potential respecting this symmetry was presented in \cite{Ivanov:2015mwl}.
It can be written as
$V = V_0+V_1$, where
\begin{eqnarray}
	V_0 &=& - m_{11}^2 (\phi_1^\dagger \phi_1) - m_{22}^2 (\phi_2^\dagger \phi_2 + \phi_3^\dagger \phi_3)
	+ \lambda_1 (\phi_1^\dagger \phi_1)^2 + \lambda_2 \left[(\phi_2^\dagger \phi_2)^2 + (\phi_3^\dagger \phi_3)^2\right]
	\nonumber\\
	&+& \lambda_3 (\phi_1^\dagger \phi_1) (\phi_2^\dagger \phi_2 + \phi_3^\dagger \phi_3)
	+ \lambda'_3 (\phi_2^\dagger \phi_2) (\phi_3^\dagger \phi_3)\nonumber\\
	&+& \lambda_4 \left[(\phi_1^\dagger \phi_2)(\phi_2^\dagger \phi_1) + (\phi_1^\dagger \phi_3)(\phi_3^\dagger \phi_1)\right]
	+ \lambda'_4 (\phi_2^\dagger \phi_3)(\phi_3^\dagger \phi_2)\,,
	\label{V0}
\end{eqnarray}
with all parameters real, and
\begin{equation}
	V_1 = \lambda_5 (\phi_3^\dagger\phi_1)(\phi_2^\dagger\phi_1) +
	\lambda_8(\phi_2^\dagger \phi_3)^2 + \lambda_9(\phi_2^\dagger\phi_3)(\phi_2^\dagger\phi_2-\phi_3^\dagger\phi_3) + h.c.
	\label{V1}
\end{equation}
with real $\lambda_5$ and complex $\lambda_8$, $\lambda_9$.
Out of the two remaining complex parameters, only one can be made real by a suitable rephasing of $\phi_2$ and $\phi_3$.
However, in this work we prefer to keep $\lambda_8$, $\lambda_9$ complex
because we will rely on the residual rephasing freedom
to make the vacuum expectation values (vevs) real.

Minimization of this potential and the resulting scalar bosons mass matrices were studied in \cite{Ferreira:2017tvy}.
In order to avoid pathologies in the quark sector, CP4 must be spontaneously broken.
The three doublets acquire vevs, which can in principle be complex,
but the rephasing freedom allows us to render the three vevs real:
\begin{equation}
	\lr{\phi_i^0} = \frac{1}{\sqrt{2}}(v_1,\, v_2,\, v_3)
	\equiv \frac{v}{\sqrt{2}}\, (c_\beta, s_\beta c_\psi, s_\beta s_\psi)\,.\label{vevs}
\end{equation}
With $v = 246$ GeV fixed, the position of the minimum is described by two angles $\beta$ and $\psi$
(in the expression above, we used the shorthand notation for sines and cosines of these angles).
Using these angles as input parameters, we rotate the three doublets to a Higgs basis as
\begin{equation}
	\triplet{H_1}{H_2}{H_3} =
	\mmmatrix{c_\beta}{s_\beta c_\psi}{s_\beta s_\psi}{-s_\beta}{c_\beta c_\psi}{c_\beta s_\psi}{0}{-s_\psi}{c_\psi}
	\triplet{\phi_1}{\phi_2}{\phi_3},.
	\label{matrix-P}
\end{equation}
In this Higgs basis, the vev is located only in $H_1$, while $\lr{H_2^0} = \lr{H_3^0} = 0$.
The would-be Goldstone modes populate $H_1$, while all fields in the doublets $H_2$, $H_3$ are physical scalars degrees of freedom.
Expanding the potential near the minimum, we obtain five neutral scalar bosons and two pairs of charged Higgses.
In general, all neutral Higgs bosons can couple to $WW$ and $ZZ$ pairs.
However, if one fixes $m_{11}^2 = m_{22}^2$, the model exhibits scalar alignment: one of the neutral Higgses
$h_{SM}$ couples to the $WW$ and $ZZ$ exactly as in the SM,
while the other four neutral bosons $h_2$ through $h_5$ decouple from these channels.
In this study, we assume scalar alignment.


\subsection{The Yukawa sector of the CP4 3HDM}

The general expressions for the quark Yukawa sector in the 3HDM is
\begin{equation}
	-{\cal L}_Y = \bar{Q}^0_L (\Gamma_1 \phi_1 + \Gamma_2 \phi_2 + \Gamma_3 \phi_3) d_R^0 +
	\bar{Q}^0_L (\Delta_1 \tilde\phi_1 + \Delta_2 \tilde\phi_2 + \Delta_3 \tilde\phi_3) u_R^0 + h.c.\label{Yukawa-general}
\end{equation}
Here, we use the notation of \cite{Botella:2018gzy} extended to the 3HDM.
The three generations of quarks are implicitly assumed everywhere, their indices suppressed for brevity.
The superscript $0$ for the quark fields indicates that these are the starting quark fields;
when we pass to the physical quark fields by diagonalizing the quark mass matrices, we will remove this superscript.

CP4 symmetry can be extended to the Yukawa sector of 3HDM.
This extension is not unique, but there is a limited number of structurally different options.
This problem was solved in \cite{Ferreira:2017tvy} and yielded four distinct cases, labeled $A$, $B_1$, $B_2$, and $B_3$,
separately in the up and down-quark sectors.
In the down-quark sector, case $A$ of \cite{Ferreira:2017tvy} represents the trivial solution
\begin{equation}
	\Gamma_1 = \mmmatrix{g_{11}}{g_{12}}{g_{13}}%
	{g_{12}^*}{g_{11}^*}{g_{13}^*}%
	{g_{31}}{g_{31}^*}{g_{33}}\,,\quad
	\Gamma_{2,3} = 0\,,\label{caseA}
\end{equation}
which is completely free from FCNCs, while 
cases $B_1$, $B_2$, $B_3$ involve all three Yukawa matrices:
\begin{itemize}
	\item
	Case $B_1$:
	\begin{equation}
		\Gamma_1 = \mmmatrix{0}{0}{0}{0}{0}{0}{g_{31}}{g_{31}^*}{g_{33}}\,,\quad
		\Gamma_2 = \mmmatrix{g_{11}}{g_{12}}{g_{13}}{g_{21}}{g_{22}}{g_{23}}{0}{0}{0}\,,\quad
		\Gamma_3 =  \mmmatrix{-g_{22}^*}{-g_{21}^*}{-g_{23}^*}{g_{12}^*}{g_{11}^*}{g_{13}^*}{0}{0}{0}\,.
		\label{caseB1}
	\end{equation}
	
	\item
	Case $B_2$:
	\begin{equation}
		\Gamma_1 = \mmmatrix{0}{0}{g_{13}}{0}{0}{g_{13}^*}{0}{0}{g_{33}}\,,\quad
		\Gamma_2 = \mmmatrix{g_{11}}{g_{12}}{0}{g_{21}}{g_{22}}{0}{g_{31}}{g_{32}}{0}\,,\quad
		\Gamma_3 =  \mmmatrix{g_{22}^*}{-g_{21}^*}{0}{g_{12}^*}{-g_{11}^*}{0}{g_{32}^*}{-g_{31}^*}{0}\,.
		\label{caseB2}
	\end{equation}
	
	\item
	Case $B_3$:
	\begin{equation}
		\Gamma_1 = \mmmatrix{g_{11}}{g_{12}}{0}{-g_{12}^* }{g_{11}^*}{0}{0}{0}{g_{33}}\,,\quad
		\Gamma_2 = \mmmatrix{0}{0}{g_{13}}{0}{0}{g_{23}}{g_{31}}{g_{32}}{0}\,,\quad
		\Gamma_3 = \mmmatrix{0}{0}{-g_{23}^*}{0}{0}{g_{13}^*}{g_{32}^*}{-g_{31}^*}{0}\,.
		\label{caseB3}
	\end{equation}
\end{itemize}
In all cases, all the parameters apart from $g_{33}$ can be complex.
A similar set of cases is found in the up-quark sector.

Since the up and down sectors involve the same left-handed doublets,
one can only combine cases $A$ or $B_2$ in the up sector with $A$ or $B_2$ in the down sectors,
and cases $B_1$ or $B_3$ in the up sector with $B_1$ or $B_3$ in the down sector.
This leads to eight possible pairings.
One of them, $(A,A)$, should be disregarded as it is unable to generate the $CP$-violating phase in the CKM matrix.
Thus, we are left with seven viable combinations for the CP4 invariant Yukawa sector:
\begin{eqnarray}
	\mbox{(down, up):} && (B_1, B_1)\,, \quad (B_1, B_3)\,, \quad (B_3, B_1)\,, \quad (B_3, B_3)\,, \label{cases-first-group}\\
	&& (A, B_2)\,, \quad (B_2, A)\,, \quad (B_2, B_2)\,. \label{cases-second-group}
\end{eqnarray}
Since case $(A,A)$ is excluded, we arrive at the important conclusion that FCNCs are unavoidable in CP4 3HDM.

\section{Controlling FCNC in the CP4 3HDM}\label{section-controlling}

\subsection{General FCNC matrices in 3HDM}

Starting from the general Yukawa sector \eqref{Yukawa-general} and inserting the vevs \eqref{vevs},
which are real in the basis we work in, we write the quark mass matrices as
\begin{equation}
	M_d^0 = \frac{v}{\sqrt{2}}(\Gamma_1 c_\beta + \Gamma_2 s_\beta c_\psi + \Gamma_3s_\beta s_\psi)\,,
	\quad
	M_u^0 = \frac{v}{\sqrt{2}}(\Delta_1 c_\beta + \Delta_2 s_\beta c_\psi + \Delta_3s_\beta s_\psi)\,.
	\label{Md0Mu0-general}
\end{equation}
They are, in general, non-diagonal and complex.
The interaction of the neutral (complex) scalars with the quarks can be described
both in the initial basis and in the Higgs basis we chose; the relation between the two bases is given by
\begin{equation}
	\Gamma_1 \phi_1^0 + \Gamma_2 \phi_2^0 + \Gamma_3 \phi_3^0 =
	\frac{\sqrt{2}}{v} (H_1^0 M_d^0 + H_2^0 N_{d2}^0 + H_3^0 N_{d3}^0)\,,
\end{equation}
where
\begin{eqnarray}
	N_{d2}^0 = M_d^0 \cot\beta -\frac{v}{\sqrt{2} s_\beta} \Gamma_1\,, \quad 
	N_{d3}^0 = \frac{v}{\sqrt{2}}(-\Gamma_2 s_\psi + \Gamma_3 c_\psi)\,.\label{Nd20Nd30-general}
\end{eqnarray}
For the up-quark sector, we obtain
\begin{equation}
	\Delta_1 (\phi_1^0)^* + \Gamma_2 (\phi_2^0)^* + \Delta_3 (\phi_3^0)^* =
	\frac{\sqrt{2}}{v} [(H_1^0)^* M_u^0 + (H_2^0)^* N_{u2}^0 + (H_3^0)^* N_{u3}^0]\,,
\end{equation}
where
\begin{eqnarray}
	N_{u2}^0 = M_u^0 \cot\beta - \frac{v}{\sqrt{2} s_\beta} \Delta_1\,,\quad
	N_{u3}^0 = \frac{v}{\sqrt{2}}(-\Delta_2 s_\psi + \Delta_3 c_\psi)\,.
\end{eqnarray}
As usual, these mass matrices are diagonalized by unitary transformations of the quark fields,
$d_L^0 = V_{dL} d_L$ and so on, which lead to the CKM matrix $V_{\rm CKM} = V_{uL}^\dagger V_{dL}$ and
\begin{equation}
	D_d = V_{dL}^\dagger M_d^0 V_{dR} = {\rm diag}(m_d, m_s, m_b)\,,\quad
	D_u = V_{uL}^\dagger M_u^0 V_{uR} = {\rm diag}(m_u, m_c, m_t)\,.
\end{equation}
The same quark rotation matrices also act on the matrices $N$:
$N_{d2} = V_{dL}^\dagger N_{d2}^0 V_{dR}$ and $N_{u2} = V_{uL}^\dagger N_{u2}^0 V_{uR}$,
and, similarly, for $N_{d3}$, $N_{u3}$.

The four Yukawa matrices $N_{d2}$, $N_{d3}$, $N_{u2}$, $N_{u3}$ are the key objects of this work.
They describe the coupling patterns
of the neutral complex fields $H_2^0$ and $H_3^0$ with the three generations of physical quarks.
Their off-diagonal elements indicate the strength of FCNC.
Within the 2HDM, we would only get $N_{d2}$ and $N_{u2}$, see e.g. \cite{Botella:2018gzy}.
The additional matrices $N_{d3}$ and $N_{u3}$ arise in the 3HDM,
and their patterns can be very different from $N_{d2}$ and $N_{u2}$.


\subsection{The inversion procedure}\label{subsection-inversion}

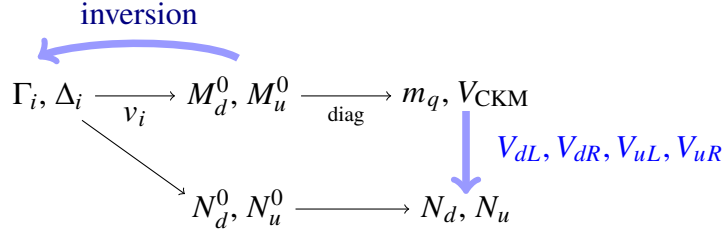
\begin{figure}[!h]
	\begin{center}
		\begin{tikzpicture}
			\large
			\node (GD) at (-0.5,1) {$\Gamma_i$, $\Delta_i$};
			\node (M0) at (2,1) {$M_d^0$, $M_u^0$};
			\node (M) at (5,1) {$m_q$, $V_{\rm CKM}$};
			\node (N0) at (2,-0.5) {$N_d^0$, $N_u^0$};
			\node (N) at (5,-0.5) {$N_d$, $N_u$};
			\draw[->] (GD) -- node[below] {$v_i$} (M0); \draw[->] (M0) -- node[below] {\scriptsize diag} (M);
			\draw[->] (GD) -- (1.3, -0.3);
			\draw[->] (N0) -- (N);
			\draw[white!60!blue,line width=1mm,->] (2,1.5) arc [start angle=20, end angle=160, x radius=1.4, y radius=.3];
			\draw[white!60!blue,line width=1mm,->] (5,0.8) -- (5,-0.3);
			\node[black!50!blue] at (0.7,2.1) {inversion};
			\node[blue] at (6.9,0.3) {$V_{dL}, V_{dR}, V_{uL}, V_{uR}$};
		\end{tikzpicture}
		\caption{\label{Fig:diagram} In a generic multi-Higgs model, one begins with $\Gamma_i$, $\Delta_i$,
			computes the mass matrices and diagonalizes them. This procedure, indicated by thin arrows, is usually irreversible.
			In certain models, one can perform inversion (thick light blue arrows),
			which allows one to pass directly from the quark properties to the FCNC matrices.}
	\end{center}
\end{figure}

Fitting quark masses and the CKM matrix is, in general,
a non-trivial task for multi-Higgs-doublet models with generic Yukawa sectors.
One begins with several Yukawa matrices $\Gamma_i$ and $\Delta_i$, which usually have many free parameters,
multiply them by vevs $v_i$ and sum them to produce the mass matrices $M_d^0$ and $M_u^0$, see Eqs.~\eqref{Md0Mu0-general}.
In general, this passage is irreversible:
if one only knows $M_d^0$ and vevs, one cannot recover individual $\Gamma_i$.
Another problem is that, if matrices $\Gamma_i$, $\Delta_i$ are not generic, 
it is not guaranteed that they can reproduce the known quark masses and mixing parameters at all, especially
when the vev alignment is also constrained by the scalar sector.

In the light of these difficulties, one is often forced to scan the multi-dimensional parameter space
in a way which is intrinsically inefficient. If one knows vevs and randomly selects $\Gamma_i$, $\Delta_i$,
one obtains $M_d^0$ and $M_u^0$, which lead to quark masses and mixing very different
from their experimental values. One then repeats the scan many times,
trying to iteratively approach the measured values.

However, in certain classes of models one can invert the above procedure:
that is, knowing $M_d^0$ and vevs, one can uniquely reconstruct each $\Gamma_i$.
%
Inversion significantly facilitates the phenomenological study of the model.
Instead of a random scan over $\Gamma_i$, $\Delta_i$, one takes the physical quark masses and mixing parameters as input,
parametrizes $V_{dL}$, $V_{dR}$, and $V_{uR}$ in a suitable way, and directly obtains a parameter space point
which automatically agrees with the experimental quark properties.
No parameter sets are wasted anymore.
Moreover, one can express the physical quark coupling matrices $N_d$ and $N_u$ via
quark masses and mixing as well as $V_{dL}$, $V_{dR}$, and $V_{uR}$.
Such a procedure offers a better control over FCNC within a given class of models.

The inversion procedure is available not only in models with natural flavor conservation, such as the Type-I or Type-II 2HDMs,
but also in the BGL model \cite{Branco:1996bq},
which allows for small FCNCs controlled by the third row of $V_{\rm CKM}$.
A similar inversion procedure was recently constructed in the $U(1)\times \Z_2$-symmetric 3HDM \cite{Das:2021oik},
where the quark sector closely resembled the BGL model.


The first scan of the CP4 3HDM parameter space reported in \cite{Ferreira:2017tvy} was done in the traditional way,
by randomly choosing $\Gamma_i$, $\Delta_i$. Although the possibility of inversion was already mentioned
in that work, it was not used in the scan. Here we report that in the recent paper \cite{Zhao:2023hws},
we successfully constructed the inversion procedure for all Yukawa sectors of the CP4 3HDM.

Then, starting from the physical quark parameters $m_q$, $V_{\rm CKM}$ and parametrizing the quark rotation matrices,
we can then calculate $N_{d2}$, $N_{d3}$, $N_{u2}$, $N_{u3}$ for the physical quark couplings
with the second and third doublet scalars $H_2^0$ and $H_3^0$ in the Higgs basis.
Although these scalars are not yet the mass eigenstates, the matrices provide a clear picture
of the FCNC magnitude and patterns.


\subsection{Target values for the FCNC magnitude}\label{subsection-target}

The Higgs-mediated contributions to the neutral meson oscillation parameters depend
on the off-diagonal elements of $N_d$ and $N_u$, on the masses of the new Higgs bosons,
and on the interference patterns among different scalars \cite{Botella:2014ska,Nebot:2015wsa}.
In this work, we address a concrete question:
is it possible to achieve, within CP4 3HDM, sufficiently small FCNC couplings 
which would satisfy all the neutral meson oscillation constraints for a 1 TeV Higgs boson 
without relying on additional cancellation?

If for some Yukawa sector of the CP4 3HDM it turns out possible,
there are good chances that a full phenomenological study
will identify viable points with reasonably heavy Higgs bosons.
Conversely, if it turns out impossible already for 1 TeV Higgses,
the chances to find a phenomenologically acceptable version 
of the explicitly CP4-invariant 3HDM will be bleak. 

To define the target parameters, we follow \cite{Nebot:2015wsa}
and rewrite the coupling matrix of a generic real scalar $S$ 
of unspecified $CP$ properties with down quarks as
\begin{equation}\label{A-B}
\frac{1}{v}\bar{d}_{Li}\, (N_d)_{ij} d_{Rj} + h.c =
\bar d_{i}\left(A_{ij} + i B_{ij} \gamma^5\right) d_j\,, \quad
	A = \frac{N_d + N_d^\dagger}{2v}\,, \quad
	i B = \frac{N_d - N_d^\dagger}{2v}\,.
\end{equation}
Both $A$ and $B$ are hermitean matrices.
For example, $K^0$--$\overline{K^0}$ oscillations place constraints on
$|a_{ds}| = |A_{12}|$ and $|b_{ds}| = |B_{12}|$, and so on.
A similar construction for the up-quark sector allows us to constrain the $(uc)$ elements
with the aid of $D^0$--$\overline{D^0}$ oscillations.
We consider the off-diagonal FCNC elements acceptable for a 1 TeV scalar
if they satisfy the following upper limits borrowed from \cite{Nebot:2015wsa}:
\begin{subequations}\label{oscillation-constraints}
	\begin{align}
		K^0 - \overline{K^0}: &\qquad |a_{ds}| < 3.7\times 10^{-4}\,, \quad |b_{ds}| < 1.1\times 10^{-4}\,,\label{constraint-ds}\\
		B^0 - \overline{B^0}: &\qquad |a_{db}| < 9.0\times 10^{-4}\,, \quad |b_{db}| < 3.4\times 10^{-4}\,,\label{constraint-db}\\
		B_s^0 - \overline{B_s^0}: &\qquad |a_{sb}| < 45\times 10^{-4}\,, \quad\  |b_{sb}| < 17\times 10^{-4}\,,\label{constraint-sb}\\
		D^0 - \overline{D^0}: &\qquad |a_{uc}| < 5.0\times 10^{-4}\,, \quad |b_{uc}| < 1.8\times 10^{-4}\,.\label{constraint-uc}
	\end{align}
\end{subequations}
For a lower mass of the scalar $S$, the upper limits on these couplings will decrease proportionally.


Below, we will check one by one all seven Yukawa sectors given 
in Eqs.~\eqref{cases-first-group} or \eqref{cases-second-group}.
If a sector manages to produce ``viable'' points, that is, parameter sets
which pass all the above constraints, we consider this case promising.
If a sector is unable to produce points passing simultaneously all the constraints,
we say the this sector is ``ruled out''.


\section{FCNCs in the CP4 3HDM: results}\label{section-FCNC-cases}

\subsection{General expressions}

Using the expressions for matrices $\Gamma_i$ in Eq.~\eqref{caseB1}--\eqref{caseB3} and the vevs
parametrized via $\beta$ and $\psi$ in Eq.~\eqref{vevs}, we can explicitly compute $M_d^0$ as well as
$N_{d2}^0$ and $N_{d3}^0$. Our goal is to relate them, that is, to express $N_{d2}^0$ and $N_{d3}^0$
via $M_d^0$, which will then allow us to express $N_{d2}$ and $N_{d3}$ in terms of quark masses and quark rotation matrices.

The detailed derivation and the exact results for all the cases $B_1$, $B_2$, and $B_3$
can be found in the paper \cite{Zhao:2023hws}. Here we only illustrate the results with the case $B_1$.
We begin with $N_{d2}^0$, Eq.~\eqref{Nd20Nd30-general}, which can be expressed as
\begin{equation}
	N_{d2}^0 = M_d^0 \cot\beta - \frac{v}{\sqrt{2} s_\beta} \Gamma_1 = R_3^0 \cdot M_d^0\,.
\end{equation}
Here, $R_3^0 = \mbox{diag}(\cot\beta, \cot\beta, -\tan\beta)$.
After the mass matrix bidiagonalization, we get
\begin{equation}
	N_{d2} = V_{dL}^\dagger N_{d2}^0 V_{dR} = V_{dL}^\dagger R_3^0 V_{dL} \cdot V_{dL}^\dagger M_d^0 V_{dR} = R_3 \cdot M_d\,.
\end{equation}
This leads to a simple expression for the FCNC matrix $N_{d2}$, which makes it clear that
the off-diagonal elements of $N_{d2}$ are fully controlled by the third row of the matrix $V_{dL}$:
\begin{equation}
	(N_{d2})_{ij} = \cot\beta\, m_{d_j}\delta_{ij} - \frac{m_{d_j}}{c_\beta s_\beta} (V_{dL, 3i})^* V_{dL, 3j}\,.\label{B1-Nd2-general}
\end{equation}
A similar analysis holds for the up-quark sector. 
These expressions are familiar from the BGL model and the $U(1)\times \Z_2$-symmetric 3HDM studied in \cite{Das:2021oik}.
This is not surprising: with our definition of the Higgs basis, the second doublet $H_2$ of our model
matches the second doublet of the 2HDM in the Higgs basis.

$N_{d3}$ has no counterpart in the BGL model. It turns out that, thanks to the special form of the Yukawa matrices 
in CP4 3HDM, the matrix $N_{d3}^0$ can be expressed in terms of $M_d^{0 *}$:
\begin{equation}
	N_{d3}^0 = \frac{1}{s_\beta} P_4 \cdot M_d^{0 *} \cdot R_2\,, \quad
	\mbox{where}\quad P_4 = \mmmatrix{0}{-1}{0}{1}{0}{0}{0}{0}{0}\,, \quad
	R_2 = \mmmatrix{0}{1}{0}{1}{0}{0}{0}{0}{1}\,. \label{B1-Nd30-general}
\end{equation}
After the quark field rotations, we get
\begin{equation}
	N_{d3} = 
	\frac{1}{s_\beta} V_{dL}^\dagger P_4 V_{dL}^*\cdot V_{dL}^T M_d^{0 *}  V_{dR}^*\cdot V_{dR}^T R_2 V_{dR}
	= \frac{1}{s_\beta} P_4^{(dL)} \cdot M_d \cdot R_2^{(dR)}\,. \label{B1-Nd3-general}
\end{equation}
Here, we used the fact that the diagonal matrix $M_d$ is real. We also defined
\begin{equation}
	P_4^{(dL)} = V_{dL}^\dagger P_4 V_{dL}^*\,, \quad
	R_2^{(dR)} = V_{dR}^T \, R_2 \, V_{dR}\,.\label{P4R2}
\end{equation}
Using the similar approach, one can express $N_d$'s and $N_u$'s
for the remaining cased $B_2$ and $B_3$, see details in \cite{Zhao:2023hws}.

A qualitative analysis presented in \cite{Zhao:2023hws} allows one to understand 
how the parameters of the quark rotation matrices affect the magnitude of the 
off-diagonal elements. The main qualitative observations are that, first,
it is easy to suppress the FCNC couplings from $N_{d2}$ but not from $N_{d3}$,
and second, that the $(ds)$ off-diagonal couplings from $N_{d3}$, which are relevant for kaon mixing, 
are governed by $m_s$,
not $m_d$ or $\sqrt{m_dm_s}$. This is due to the fact that the CP4 symmetry mixes 
the first two quark generations.
Similarly, the $D$-meson mixing parameters receive contributions from $(uc)$ couplings
of the matrix $N_{u3}$, which inherit $m_c$.
All these off-diagonal couplings can still be suppressed by the small rotation angles,
but the presence of the second generation quark masses explains why it is a highly non-trivial task
to avoid large FCNC in the CP4 3HDM.

\subsection{Numerical results}

We are now ready to address the main question formulated in Section~\ref{subsection-target}:
is it possible to find examples of the CP4 3HDM Yukawa sectors
which satisfy all the meson oscillation constraints given in Eqs.~\eqref{oscillation-constraints}?
We implemented the inversion procedure
described in Section~\ref{section-controlling}
using the experimentally known quark masses, the CKM matrix, and the scalar vevs as input parameters.
We chose the quark rotation matrices in such a way that, in the original basis
before the quark mass matrices diagonalization, we could recover $M_d^0$ and $M_u^0$
exactly of the type that is required for each Yukawa sector.
For cases $A$, $B_1$, and $B_2$, this can be done analytically,
while for case $B_3$ we resorted to a numerical procedure.

The angles and phases of the quark rotation matrices represent the space
in which we performed numerical scans.
We ran different versions of the scan, either exploring the full ranges of the rotation angles and phases
or staying close to the block-diagonal quark rotation matrices.
For each parameter space point, we computed the dimensionless FCNC quantifies $a_{ij}$ and $b_{ij}$
defined in Section~\ref{subsection-target} and checked whether they could satisfy the constraints
at least for a 1 TeV scalar.

\begin{figure}[!h]
	\centering
	\includegraphics[width=0.48\textwidth]{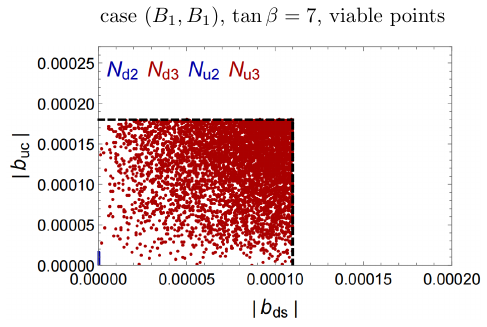}
	\includegraphics[width=0.48\textwidth]{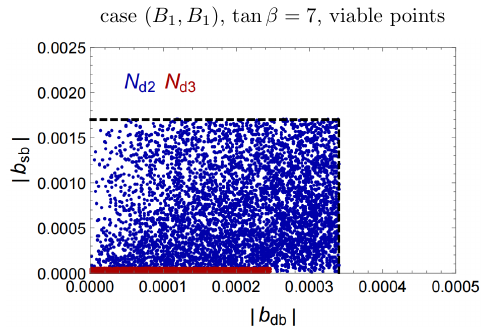}
	\caption{Constraints from the kaon and $D$-meson oscillations (left) and from $B$/$B_s$-meson oscillations (right)
		for the subset of points which satisfy all the meson oscillation constraints.
		The dashed boxes show the constraints of Eq.~\eqref{oscillation-constraints}.
	}
	\label{fig-B1B1-good}
\end{figure}

Starting with the CP4-invariant Yukawa scenario $(B_1,B_1)$, we observed the off-diagonal entries of 
the Higgs-quark coupling matrices $N_{d2}$, $N_{d3}$, $N_{u2}$, $N_{u3}$
could be controlled through appropriate parameters of the quark rotation matrices,
especially by choosing the quark rotation matrices sufficiently close to the block-diagonal form. 
We also noticed that the $D$-meson oscillations placed the strongest constraints. 
The fact that the previous numeriucal analysis of CP4 3HDM \cite{Ferreira:2017tvy} did not include $D$-mesons
makes it well possible that all the $(B_1,B_1)$ points considered there as viable could
be in fact ruled out by the $D$-meson constraints. 

In Fig.~\ref{fig-B1B1-good} we show viable points, which pass all the meson oscillation constraints,
on the $D$-meson vs kaon FCNC coefficients plane (left)
and on the $B_s$ vs $B$-meson coefficients plane (right). 
These plots show that couplings of $H_2^0$ and $H_3^0$ 
to quarks are shaped by different meson oscillation constraints. 
$N_{d2}$ and $N_{u2}$ (blue points) satisfy the kaon and $D$-meson constraints by large margin,
and their main limitations come from $B$ physics.
For $N_{d3}$ and $N_{u3}$ (red points), we observe the opposite trend: $B$ physics constraints play a minor role,
and the strongest restrictions come from kaons and $D$-mesons.
This is a manifestation of the structurally different forms of the corresponding matrices.

\begin{figure}[t]
	\centering
	\includegraphics[width=0.6\textwidth]{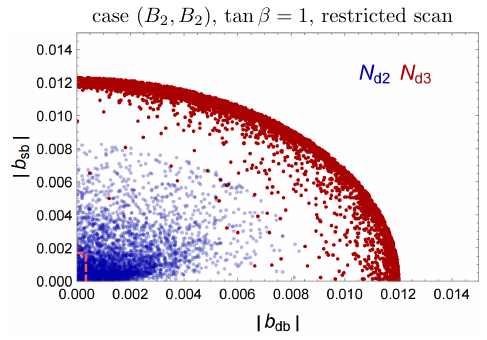}
	\caption{Constraints from the $B$/$B_s$-meson oscillations
		on the CP4-invariant Yukawa case $(B_2,B_2)$. All red points fall outside the box, shown by the red dashed line.}
	\label{fig-B2B2}
\end{figure}

This first example shows that the Yukawa sector $(B_1,B_1)$ is a promising option for a CP4 3HDM benchmark model.
It turns out that in other Yukawa scenarios of the CP4 3HDM, 
significant tension arises among several meson oscillation constraints.
For an illustration, we show in Fig.~\ref{fig-B2B2} the scan results for the Yukawa case $(B_2,B_2)$
and show the distribution of $b_{db}$ and $b_{sb}$, which are the most tightly constrained
coefficients affecting $B$ and $B_s$ meson oscillations.
Although off-diagonal elements of the matrix $N_{d2}$ (blue points), describing the coupling of $H_2^0$ to quarks,
can be well suppressed, the corresponding elements of the $H_3^0$ coupling matrix $N_{d3}$ (red points)
are outside the box far large margin. This means that, in a full phenomenological study, 
the Yukawa scenario $(B_2,B_2)$ would have a chance to satisfy all the constraints only 
if we used the new scalar masses of the order of several TeV. But the scalar sector of the CP4 3HDM model 
does not have decoupling limit, so such high masses can never be attained.
Thus, the scenario $(B_2,B_2)$ is already ruled out for CP4 3HDM.

Apart from $(B_1,B_1)$, only the Yukawa scenario $(A,B_2)$ can produce viable points.
In this case, the down quark sector has no FCNC at all, and the only constraint 
comes from the $D$-meson oscillations. We found that, in a certain region of the parameter space, 
it could be satisfied simultaneously for $N_{u2}$ and $N_{u3}$.

\section{Summary}

Neutral meson oscillations put to a severe test all multi-Higgs models
that feature Higgs-induced tree-level flavor changing neutral couplings \cite{Sher:2022aaa}.
CP4 3HDM, the three-Higgs-doublet model with a $CP$ symmetry of order 4, 
features strong connections between the scalar and Yukawa
sectors and, therefore, possesses potentially dangerous FCNCs.
In this comunication, we reported the results of \cite{Zhao:2023hws} in which
we investigated whether and how strongly FCNC effects be suppressed within CP4 3HDM.

To this end, we constructed a new scanning procedure which uses
the quark masses and CKM matrix as input parameters, expressed the FCNC matrices
of the new scalars via these physical parameters and the quark rotation matrices,
and performed scans over angles and phases of the rotation matrices.
Using this procedure, we found that out of eight possible CP4 3HDM Yukawa sectors,
only two could lead to potentially viable models; 
these are the scenarios $(B_1,B_1)$ and $(A,B_2)$.
All the remaining scenarios unavoidably lead to FCNC couplings which are too large
even for a 1 TeV new scalar, and we consider them ruled out.

We are now working on a full phenomenological study of the CP4 3HDM based on either of these two Yukawa scenarios.


\acknowledgments 
I.P.I. wishes to thank the organizers of the Corfu Summer Institute for inviting him
to this inspiring workshop and setting up the wonderful working atmosphere. 

\providecommand{\href}[2]{#2}\begingroup\raggedright

\end{document}